\documentclass[]{arXiv_v1}

\preprintnumber{XXXX-XXXX} 
\usepackage{hyperref}
\usepackage{comment}
\usepackage{enumitem}

\begin{document}

\title{Measurement of $\gamma$-rays generated by neutron interaction with ${}^{16}$O at 30 MeV and 250 MeV}

\author[1]{T.~Tano}
\author[1]{T.~Horai}
\author[2,8]{Y.~Ashida}
\author[1,*]{Y.~Hino} 
\author[3]{F.~Iacob}
\author[4]{A.~Maurel}
\author[5]{M.~Mori}
\author[3]{G.~Collazuol}
\author[6,7]{A.~Konaka}
\author[1]{Y.~Koshio}
\author[8]{T.~Nakaya}
\author[7]{T.~Shima}
\author[8]{R.~Wendell}

\affil[1]{Department of Physics, Okayama University, Okayama 700-8530, Japan}
\affil[2]{Department of Physics and Astronomy, University of Utah, 115 S 1400 E, Salt Lake City, USA}
\affil[3]{Department of Physics and Astronomy, University of Padova, Padova 8-35131, Italy}
\affil[4]{Ecole Polytechnique, IN2P3-CNRS,Laboratoire Leprince-Ringuet, F-91120 Palaiseau, France}
\affil[5]{National Astronomical Observatory of Japan, Mitaka, Tokyo 181-8588, Japan}
\affil[6]{TRIUMF, Vancouver V6T 2 A3, Canada}
\affil[7]{Research Center for Nuclear Physics (RCNP), Osaka 567-0047, Japan}
\affil[8]{Department of Physics, Kyoto University, Kyoto 606-8502, Japan
\email{yotahino@okayama-u.ac.jp}
}






\begin{abstract} 
Deep understanding of $\gamma$-ray production from the fast neutron reaction in water is crucial for various physics studies at large-scale water Cherenkov detectors. 
We performed test experiments using quasi-mono energetic neutron beams ($E_n = 30$ and 250~MeV) at Osaka University's Research Center for Nuclear Physics to measure $\gamma$-rays originating from the neutron-oxygen reaction with a high-purity germanium detector. 
Multiple $\gamma$-ray peaks which are expected to be from excited nuclei after the neutron-oxygen reaction were successfully observed. 
We measured the neutron beam flux using an organic liquid scintillator for the cross section measurement. 
With a spectral fitting analysis based on the tailored $\gamma$-ray signal and background templates, we measured cross sections for each observed $\gamma$-ray component. 
The results will be useful to validate neutron models employed in the on-going and future water Cherenkov experiments. 
\end{abstract}

\subjectindex{xxxx, xxx}

\maketitle

\section{Introduction}
\label{sec:intro}

\begin{figure}
    \centering
    \includegraphics[width=0.8\textwidth]{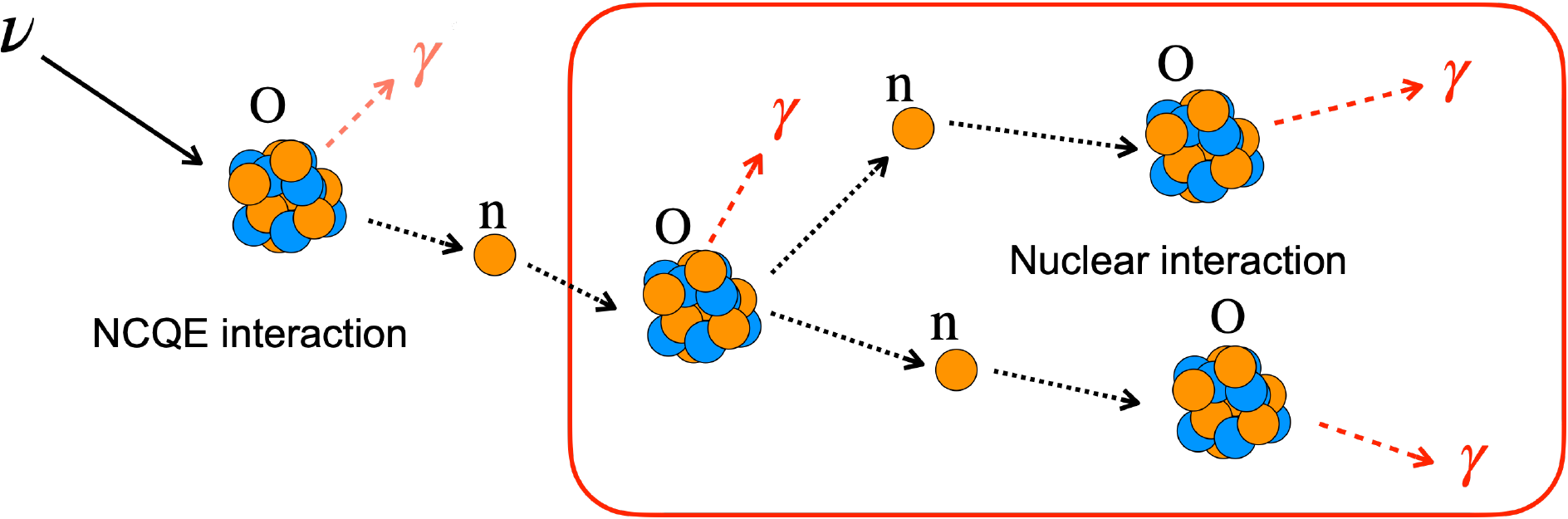}
    \caption{Schematic illustration of the neutrino-oxygen NCQE interaction and the secondary interaction by knocked-out neutrons.}
    \label{fig:schematic_ncqe}
\end{figure}

A precise modeling of the neutrino-oxygen neutral-current quasi-elastic (NCQE) interactions is essential for various studies of astrophysics and particle physics at large-scale water Cherenkov detectors, such as Super-Kamiokande (SK)~\cite{bib:superk} and its successor, Hyper-Kamiokande~\cite{bib:hyperk}. 
For example, atmospheric neutrino NCQE interactions on ${\rm ^{16}O}$ in water needs to be estimated accurately as they mimic the inverse beta decay signal ($\bar{\nu}_e + p \to e^+ + n$) of the diffuse supernova neutrino background (DSNB)~\cite{2021PhRvD.104l2002A,Super-Kamiokande:2023xup}. 
There have been multiple neutrino-oxygen NCQE cross section measurements performed by using accelerator neutrino beams in T2K~\cite{2014PhRvD..90g2012A,2019PhRvD.100k2009A} and atmospheric neutrinos at SK~\cite{2019PhRvD..99c2005W,2024PhRvD.109a1101S}. 
At neutrino energies of $E_{\nu} \gtrsim 200$~MeV, a single nucleon knock-out is expected to occur dominantly after the NCQE interaction~\cite{2012PhRvL.108e2505A,2013PhRvD..88i3004A} as, 
  \begin{equation}
    \begin{split}
      \nu\,(\bar{\nu}) + {}^{16}\mathrm{O} &\to \nu\,(\bar{\nu}) + {}^{15}\mathrm{O}^{*} + n, \\
      \nu\,(\bar{\nu}) + {}^{16}\mathrm{O} &\to \nu\,(\bar{\nu}) + {}^{15}\mathrm{N}^{*} + p, 
    \end{split}
  \end{equation}
where an excited nucleus is left in some cases that will relax with an emission of $\gamma$-ray(s), e.g., a branching fraction of $\sim$65\% is predicted for an excited state after the NCQE interaction in Ref.~\cite{2013PhRvD..88i3004A}.
Energies of the knocked-out nucleon are typically in hundreds of MeV. 
Such nucleons, particularly neutrons, would cause additional reactions inside the detector that produce more $\gamma$-rays, as schematically shown in Figure~\ref{fig:schematic_ncqe}. 
Here, the primary $\gamma$-ray from the NCQE interaction and the secondary $\gamma$-ray from the following nuclear reactions are emitted within only O(10)~ns, which makes it difficult to separate from each other, and then both $\gamma$-rays are treated as signals in analysis. 
For this reason, an accurate modeling of secondary $\gamma$-ray emission in the neutron-oxygen reaction is essential to predict the signal distribution in the NCQE measurement at water Cherenkov detectors. 
Currently, it is implied that there is an issue with modeling of the secondary $\gamma$-ray emission, causing a large systematic uncertainty in the cross section measurement~\cite{2019PhRvD.100k2009A,2024PhRvD.109a1101S}.   
In the T2K measurement~\cite{2019PhRvD.100k2009A}, more events are expected at large Cherenkov angles, where multiple $\gamma$-rays are typically reconstructed, in the Monte Carlo simulation than the observed data. 
In the SK measurement~\cite{2024PhRvD.109a1101S}, several nuclear models were tested for the NCQE sample and a potential ill-modeling of the secondary $\gamma$-ray emission is pointed out. 
Therefore, there is a high demand that more experimental data are provided for better modeling of the nuclear $\gamma$-ray production in water. 

The ${}^{16}\mathrm{O}(n, \mathrm{X}\gamma)$ cross sections have been measured for various $\gamma$-ray components using quasi-mono energetic neutron beams at Osaka University's Research Center for Nuclear Physics (RCNP). 
In order to cover the energy range of the knocked-out nucleon induced by the neutrino-oxygen NCQE interaction, neutron energies of 30, 80, and 250~MeV were chosen for the measurements. 
The first RCNP experiment was conducted in 2017 as E487 at $E_n = 80$~MeV~\cite{2024PhRvC.109a4620A}. 
Additional beam experiments, E525, were performed at $E_n = 30$ and 250~MeV, and their ${}^{16}\mathrm{O}(n, \mathrm{X}\gamma)$ cross section results are reported in this paper. 
The rest part of the paper is structured as follows. 
In Section~\ref{sec:experiment}, the RCNP facility and our experimental setup are described. 
In Sections~\ref{sec:neutron_flux} and \ref{sec:gamma_ray}, the neutron flux and $\gamma$-ray measurements are detailed. 
The ${}^{16}\mathrm{O}(n, \mathrm{X}\gamma)$ cross section results from the experiments are reported in Section~\ref{sec:cross_section}. 
Discussion about our results and impacts to the NCQE measurement at water Cherenkov detectors is given in Section~\ref{sec:discussion}, before concluding in Section~\ref{sec:conclusion}.

\section{The RCNP-E525 Experiment} \label{sec:experiment}

The E525 experiment was carried out in a 100 m long beamline at RCNP. 
The basic setup and measurement strategy followed the previous experiment (E487) performed in the same beamline~\cite{2024PhRvC.109a4620A}, while several parts have been optimized or improved. 
Protons were accelerated by two cyclotrons, the K140 azimuthally varying field cyclotron and the K400 ring cyclotron, and guided into a lithium target to produce a quasi-mono energetic neutron beam via the ${}^7{\rm Li}(p,n){}^7{\rm Be}$ reaction. 
The lithium target consists of 7.5\% ${}^6{\rm Li}$ and 92.5\% ${}^7{\rm Li}$ and its thickness is 10.0$\pm$0.2~mm.
In E525, we performed measurements at two proton energy settings, 30~MeV and 250~MeV, on December 16--17 and October 30--31 in 2018, respectively. 
The resulting beam entering the experimental hall is dominated by neutral particles as charged particles are rejected after the lithium target by a magnetic field. 
The proton beam current was being monitored by a Faraday cup placed at the beam dump and its data was read out by a scaler module. 

  \begin{figure}[htbp]
    \centering\includegraphics[width=\linewidth]{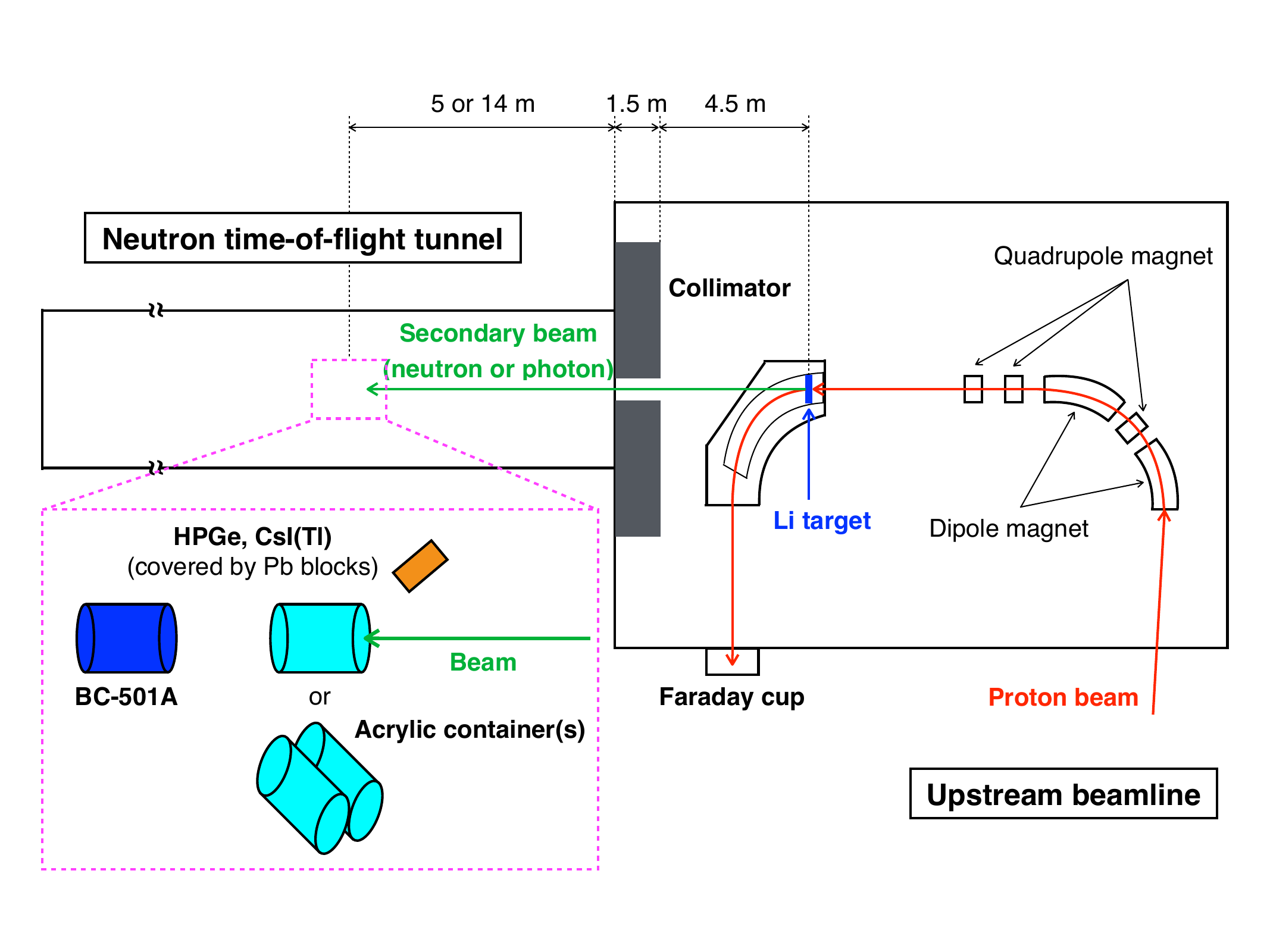}	
  \vspace{-35truept}
  \caption{Schematic drawing of the facility and beamline for the E525 experiment at RCNP. Our detectors were placed at 11 (20)~m downstream of the lithium target in the 30 (250)~MeV experiment. The collimator size is $10 \times 12$~cm.}
  \label{fig:schematic_view}
  \end{figure}

Figure~\ref{fig:schematic_view} shows a schematic view of the beamline as well as our detector setup therein. 
In the 30~MeV experiment, we placed an acrylic container which measures 20 cm in diameter and 26.5 cm in length on the beam axis 11~m away from the lithium target.
In the 250~MeV experiment, in order to measure the time-of-flight distribution accurately, we set up the water sample and detectors farther, 20~m away from the target. 
Accordingly, to gain enough data statistics, we used two identical acrylic containers and placed them as their side facing the beam axis (see Figure~\ref{fig:schematic_view}). 
A high-purity germanium detector (HPGe) and a CsI(Tl) scintillator were placed at upstream of the acrylic container(s) off the beam axis in order to avoid neutron background. 
The location was determined by looking at the neutron beam profile measured in each experiment as described in Section~\ref{sec:gamma_ray}. 
A liquid scintillator (BC-501A) was aligned at downstream of the container and on the beam axis. 
These container(s) and detectors are the same ones as the E487 experiment \cite{2024PhRvC.109a4620A}. 
All detector waveforms were recorded with a 14-bit 250-MHz CAEN DT5725 digitizer. 
Measurements were performed in two configurations with the container(s) being water-filled and empty. 

There are several updated aspects in E525 from the previous E487 experiment. 
The main $\gamma$-ray detector in E525 is HPGe whose energy resolution is much better than the ${\rm LaBr_{3}(Ce)}$ scintillator used in E487. 
This helped us to distinguish multiple $\gamma$-ray peaks more clearly.
In E487, there might be a potential issue with the proton current measurement by a Faraday cup and a scaler module. 
The proton current setting was changed frequently back-and-forth between the neutron flux measurement ($\sim$1~nA) and the $\gamma$-ray measurement ($\gtrsim$100~nA) because of the setup and this required a frequent switch of the scaler's scale setting, which might lead to a potential human error causing a wrong normalization in the $\gamma$-ray measurement. 
This will be described in more detail in Section~\ref{sec:discussion}. 
We took a special care of this issue in the 250 MeV experiment by avoiding frequent switch of the beam power. 
In the 30 MeV experiment, in addition, we monitored the neutron flux during the $\gamma$-ray measurement using the BC-501A detector placed at downstream of the water sample. 


\section{Neutron Flux} \label{sec:neutron_flux}

Neutron flux measurement was performed using a liquid scintillator, BC-501A, and repeated between the main $\gamma$-ray measurements in order to monitor the beam stability throughout the experiment. 
The water sample was removed during the flux measurement in order to avoid attenuation by water. 
In the following part, we describe each essential portion in the flux analysis: a pulse shape discrimination (PSD) technique for neutron identification, a time-of-flight (ToF) method for reconstruction of the neutron kinetic energy, and the SCINFUL-QMD Monte Carlo code \cite{bib:SCINFUL_JAEA} used for neutron detection efficiency by BC-501A.

\subsection{Neutron event selection}
\label{subsec:nflux_PSD}
\begin{figure}[h]
    \centering
    \includegraphics[width=\linewidth]{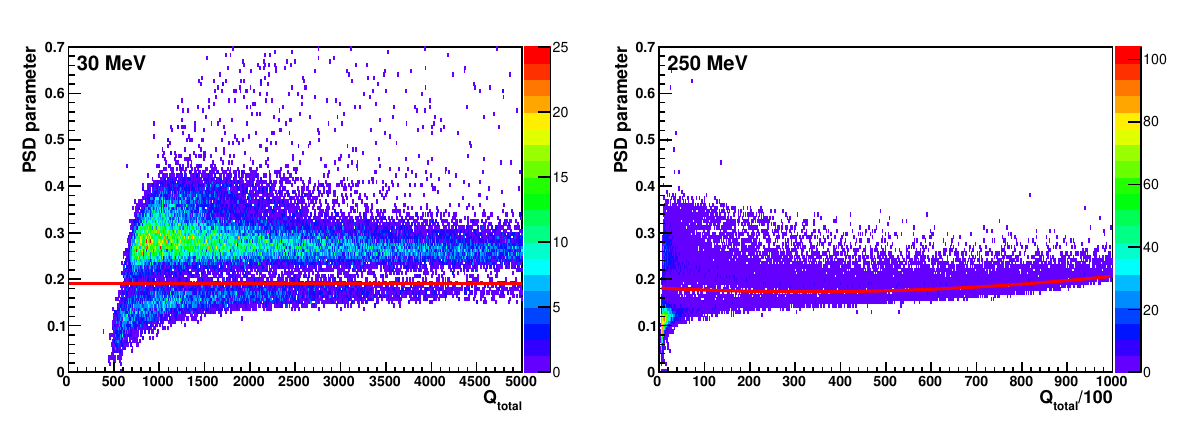}	
    \caption{
      Calculated PSD parameters as a function of $Q_{\mathrm{total}}$ for the 30~MeV (left) and 250~MeV experiment (right). The cut criteria for neutron selection are displayed in red. 
      Due to the saturation effect of electronics, a correlation between $Q_{\mathrm{total}}$ and $PSD = Q_{\mathrm{tail}}/Q_{\mathrm{total}}$ shows a non-linearity in the large charge region in the 250~MeV experiment.
    \label{fig:psd_vs_qtotal}}
\end{figure}

Scintillation light emission in an organic scintillator from nuclear recoil, e.g., a proton recoiled by neutron elastic scattering, contains a larger component with a slow time constant than electron recoil due to larger $dE/dx$.
This phenomenon appears in the shape of the resulting waveform, i.e., neutron-like events show a longer tail than $\gamma$-ray-induced events.
In order to characterize the waveform shape, in particular the fraction of the tail part, the PSD parameter is defined as,  
\begin{equation}
    PSD = \frac{Q_{\mathrm{tail}}}{Q_{\mathrm{total}}},
\end{equation}
where $Q_{\mathrm{total}}$ and $Q_{\mathrm{tail}}$ are the integrated charge for the entire waveform and its tail part, respectively. 
We optimized the integral time windows for $Q_{\mathrm{total}}$ and $Q_{\mathrm{tail}}$ as $\left[ T_0, T_0+600~\mathrm{ns} \right]$ and $\left[ T_0+48, T_0+600~\mathrm{ns} \right]$, respectively, where $T_0$ is defined as a rise time of the waveform. 
The left panel of Figure~\ref{fig:psd_vs_qtotal} shows a PSD parameter as a function of $Q_{\mathrm{total}}$ from the 30~MeV experiment. 
Here, our optimized criteria for selecting neutron-like events, $PSD > 0.19$, is also shown. 
Because the saturation effect of electronics gets more visible at high charges and was actually observed in the 250~MeV experiment, the selection curve is modified as, 
\begin{equation}
    PSD > p_0 + p_1 \times Q_{\mathrm{total}} + p_2 \times Q_{\mathrm{total}}^2.
\end{equation}
With optimization, we determined $p_0 = 0.181$, $p_1 = -5.79 \times 10^{-4}$, and $p_2 = 1.04 \times 10^{-5}$, as shown in the right panel of Figure~\ref{fig:psd_vs_qtotal}. 
The selection efficiencies of the PSD cut achieved more than 99\% in both experiments, which gives a negligibly small effect on neutron detection efficiency discussed later.
Mis-identification ratio of electron recoil event to nuclear recoil in the criterion was estimated as 0.8\% in the 30~MeV experiment and 0.5\% in the 250~MeV experiment.

\subsection{Kinetic energy reconstruction}
\label{subsec:nflux_ToF}
\begin{figure}[h]
    \centering
    \includegraphics[width=\linewidth]{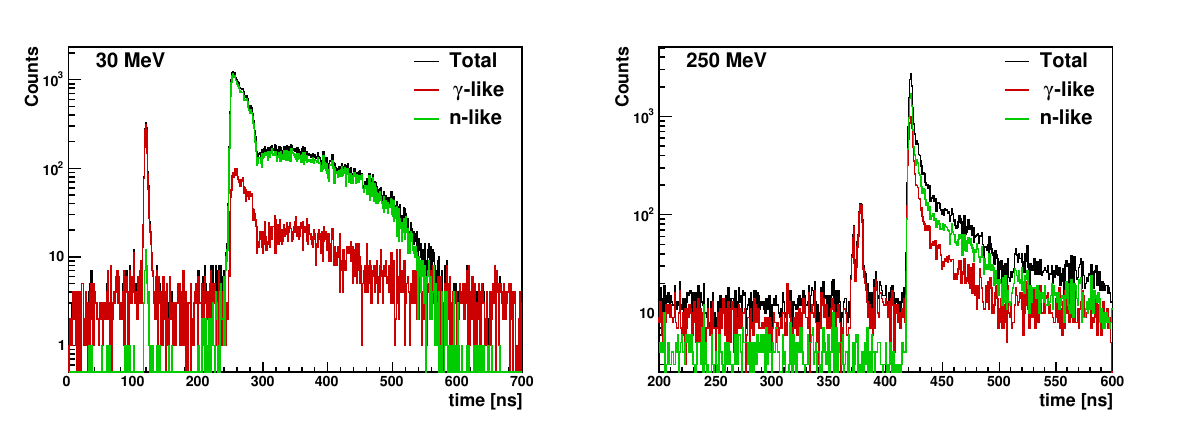}	
    \caption{ToF distributions categorized as $\gamma$-like (red), neutron-like (green) and all events (black) in the 30~MeV (left) and 250~MeV experiment (right). The peak at early times mainly consisting of the $\gamma$-like events corresponds to the flash $\gamma$-ray. The neutron from the ${}^7{\rm Li}(p,n){}^7{\rm Be}$ reaction forms the peak at late times.}
    \label{fig:tof}
\end{figure}

\begin{figure}[h]
    \centering
    \includegraphics[width=\linewidth]{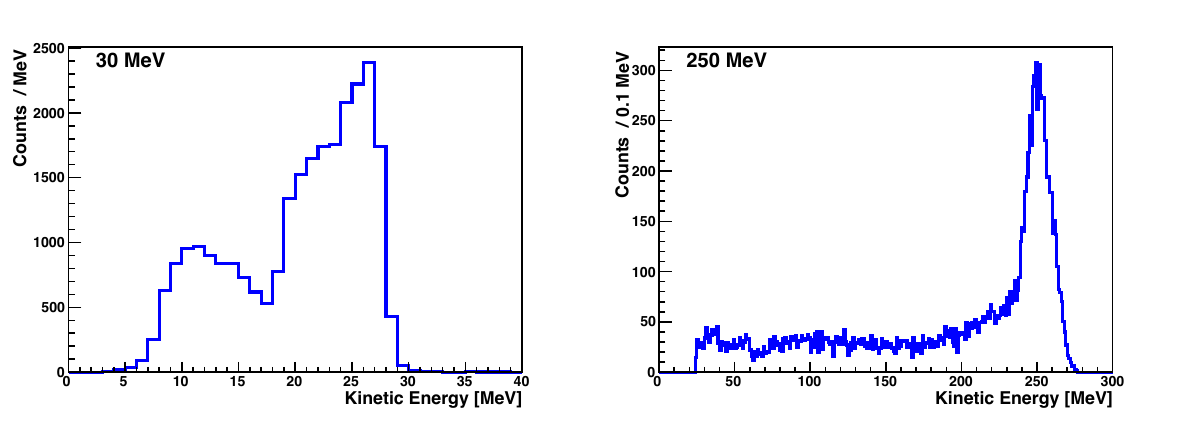}	
    \caption{Reconstructed neutron kinetic energy distributions based on ToF in the 30~MeV (left) and 250~MeV experiment (right).}
    \label{fig:neutron_energy}
\end{figure}

Neutron kinetic energies are reconstructed based on ToF, a time interval between the rise time of the BC-501A waveform and a RF-synchronized periodic pulse timing.
Figure~\ref{fig:tof} shows the ToF distributions obtained in each experiment.
The sharp peaks around 120~ns in the left panel and 370~ns in the right panel correspond to events of $\gamma$-rays instantly emitted due to the proton-lithium interaction (called ``flash $\gamma$-rays'').
One can find another peak $\sim$7~ns before the 370~ns peak in the 250~MeV experiment. 
From the detailed investigation, we confirmed that a polyimide tape inserted in the proton beam pipe for preparation of a preceding experiment caused this peak. 
The position where the polyimide tape was inserted was approximately 2.5~m upstream of the lithium target, which is consistent with the observed time difference of $\sim$7~ns. 
In addition, there were no material like the polyimide tape at the time when we performed the 30~MeV experiment.
Thus, we identified the latter peak around 370~ns as the flash $\gamma$-ray peak of the proton-lithium interaction in the 250~MeV experiment.
The resulting kinetic energy distributions converted from ToF are shown in Figure~\ref{fig:neutron_energy}. 
The peak energies, 18--30~MeV and 235--270~MeV, reconstructed in each experiment are consistent with the proton beam settings. 

\subsection{Detection efficiency \label{subsec:deteff}}
\begin{figure}[h]
    \centering
    \includegraphics[width=0.7\textwidth]{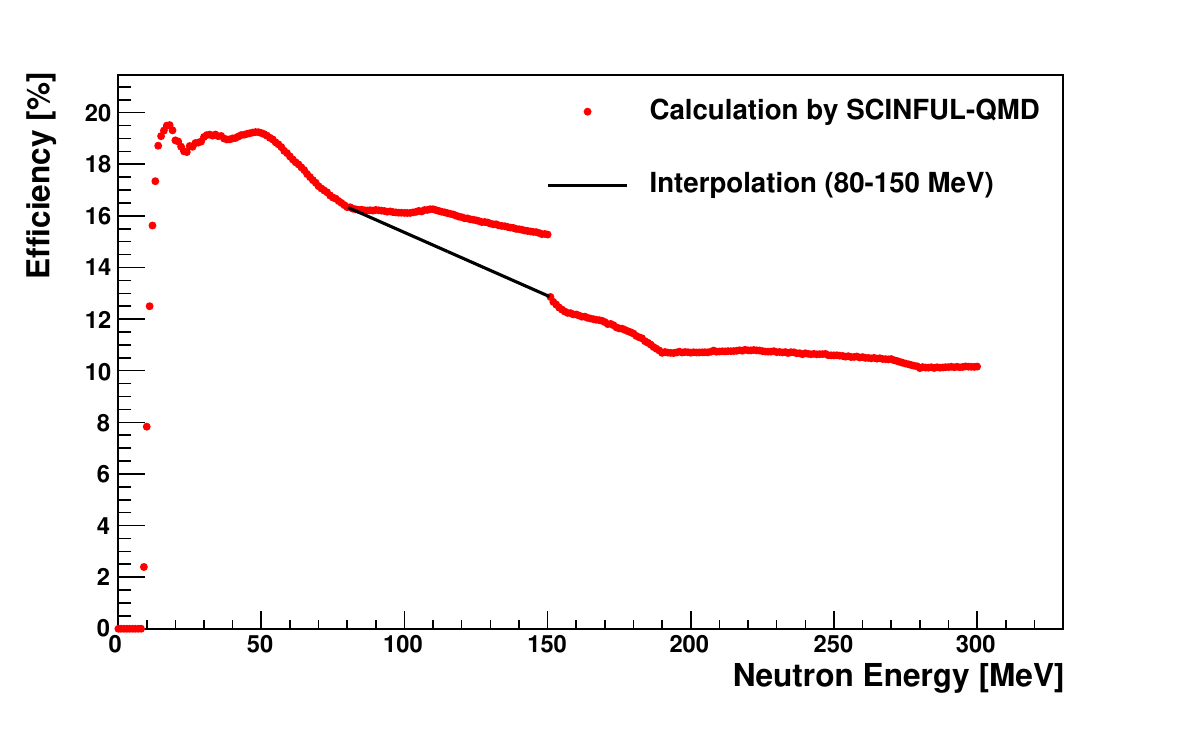}	
    \caption{Neutron detection efficiencies of the BC-501A scintillator. Red points correspond to the SCINFUL-QMD calculations and a black line gives an interpolation between 80 and 150~MeV as recommended by the developer.}
    \label{fig:detection_efficiency}
\end{figure}

A dedicated Monte Carlo code for computing neutron detection efficiency of the BC-501A scintillator, SCINFUL-QMD~\cite{bib:SCINFUL_JAEA}, is used for the efficiency calculation. 
It deals with the process of neutron reaction in organic scintillation material and conversion to PMT light yields. 
Inputs to SCINFUL-QMD include detector and source geometries, detector threshold, attenuation factor of the scintillation material and the PMT response function.
The detector threshold is confirmed using the Compton edge of the 4.4~MeV $\gamma$-ray from an ${\rm ^{241}Am/Be}$ source. 
The scintillator attenuation factor is set to 0.008 $\rm{cm}^{-1}$ following Ref.~\cite{2024PhRvC.109a4620A}.
Three functional forms are available as the PMT light output model and the difference of the resulting detection efficiency from these three models is taken into account as a systematic uncertainty in this analysis. 
Figure~\ref{fig:detection_efficiency} shows the calculated neutron detection efficiency of BC-501A as a function of neutron energy with a linear interpolation for the energy range of 80--150~MeV.
Note that because SCINFUL-QMD unnatually performs its calculation using a constant cross section value in 80 to 150~MeV, its developer recommends the interpolation prescription in this range.

\subsection{Flux estimation \label{subsec:flux_estimation}}

\begin{figure}[h]
    \centering
    \includegraphics[width=1.0\textwidth]{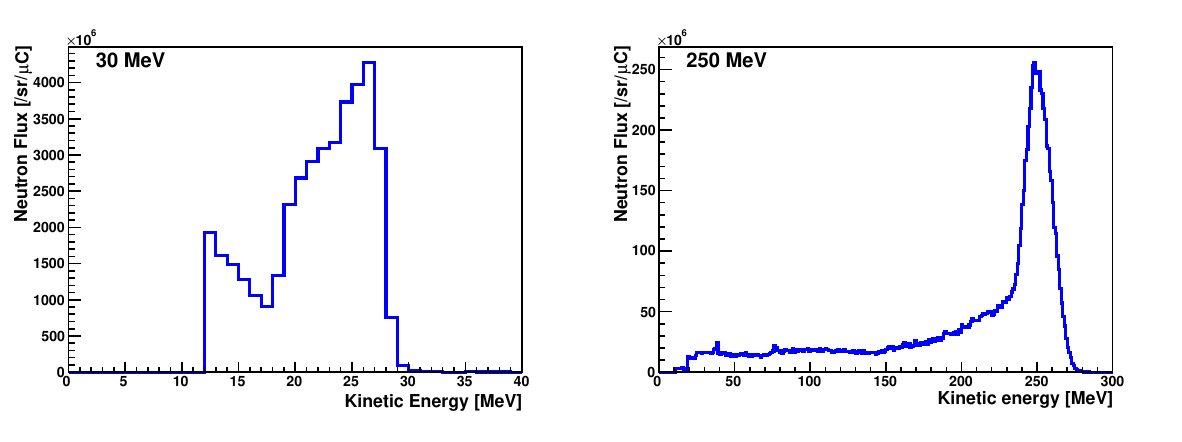}	
    \caption{Neutron flux distributions in the unit of injected proton beam current and detector solid angle in the 30~MeV experiment (left) and the 250~MeV experiment (right). The bin widths of each histogram are identical with that of Fig.~\ref{fig:neutron_energy}.}
    \label{fig:neutron_flux}
\end{figure}

Neutron flux distributions in each experiment are obtained by scaling the reconstructed neutron kinetic energy distributions with the detection efficiency and the solid angle as shown in Figure~\ref{fig:neutron_flux}.
The estimated neutron flux in the peak region is $2.11 \times 10^{10}~\rm{neutrons/sr/\mu C}$ (18--30~MeV) and $5.01 \times 10^{9}~\rm{neutrons/sr/\mu C}$ (235--270~MeV), respectively.
The flux results will be used when calculating the ${}^{16}\mathrm{O}(n, \mathrm{X}\gamma)$ cross sections. 
Table~\ref{table:flux_errors} summarizes uncertainties in the estimation of neutron flux within the 18-30 MeV and 235-270 MeV energy ranges.
The statistical uncertainties are less than 1\%.
Each source of systematic uncertainty is described below.

\subsubsection*{Detection efficiency \label{subsubsec:detection_efficiency}} \ \ 
Ref.~\cite{bib:BC-501A_attenuation} reports an uncertainty derived from physics model used in SCINFUL-QMD is 10.0\% and 15.0\% for the neutron energy below and above 80~MeV, respectively.
The effect of the SCINFUL inputs, e.g., a threshold value, light attenuation factor, choice of the light output function, and Monte Carlo statistical uncertainty, are all smaller than 0.1\%.
Consequently, 10.0\% and 15.0\% uncertainties are assinged to the detection efficiency for the 30~MeV experiment and the 250~MeV experiment, respectively.

\subsubsection*{Beam stability \label{subsubsec:beam_stability}} \ \ 
As mentioned above, the neutron flux measurements were performed between $\gamma$-ray measurements for monitoring stability. 
It was stable within 1.1\% and 10.0\% for the 30~MeV and 250~MeV experiment, respectively. They are considered as the systematic uncertainty for the neutron flux.
It is natural that there should be a larger fluctuation in the 250~MeV experiment because an electromagnet of the ring cyclotron did not reach at thermal equilibrium sufficiently. Because no other experiments were scheduled before our beam time, beam operation for the 250 MeV experiment began with cold magnet.
On the other hand, the AVF cyclotron was in operation before the 30~MeV experiment and already reached at thermal equilibrium at that time; thus, the beam was stable in the 30~MeV experiment.

\subsubsection*{Neutron selection with PSD \label{subsubsec:neutron_selection}} \ \ 
The systematic uncertainty associated with the PSD cut is estimated based on contamination of the $\gamma$-like events in the neutron-like events. 
We consider 0.8\% and 0.5\% as an uncertainty for each 30 and 250~MeV experiment, respectively.

\subsubsection*{Solid angle \label{subsubsec:solid_angle}} \ \ 
The result is 0.2\% in the 30~MeV experiment and 0.1\% in the 250~MeV experiment.

\begin{table}[h]
    \centering
    \caption{Statistical and systematic uncertainties of the neutron flux measurement.}
    \begin{tabular}{lll}
    \hline \hline
    Error Source 						& Size (18--30 MeV) [\%] & Size (235--270 MeV) [\%] \\ \hline
    Statistical 						& 1.0 					& 0.4 \\ \hline
    Detection efficiency by SCINFUL-QMD & 10.0  				& 15.0 	\\
    Beam stability 						& 1.1 					& 10.0	\\
    Neutron selection 					& 0.8 					& 0.5\\
    Solid angle 						& 0.2 					& 0.1\\ \hline
    Total 								& 10.1 					& 18.0\\ \hline \hline
    \end{tabular}
    \label{table:flux_errors}  
\end{table}


\section{Nuclear Deexcitation $\gamma$-rays} \label{sec:gamma_ray}

\subsection{Detector positioning}

\begin{figure}[h]
    \centering
    \includegraphics[width=\linewidth]{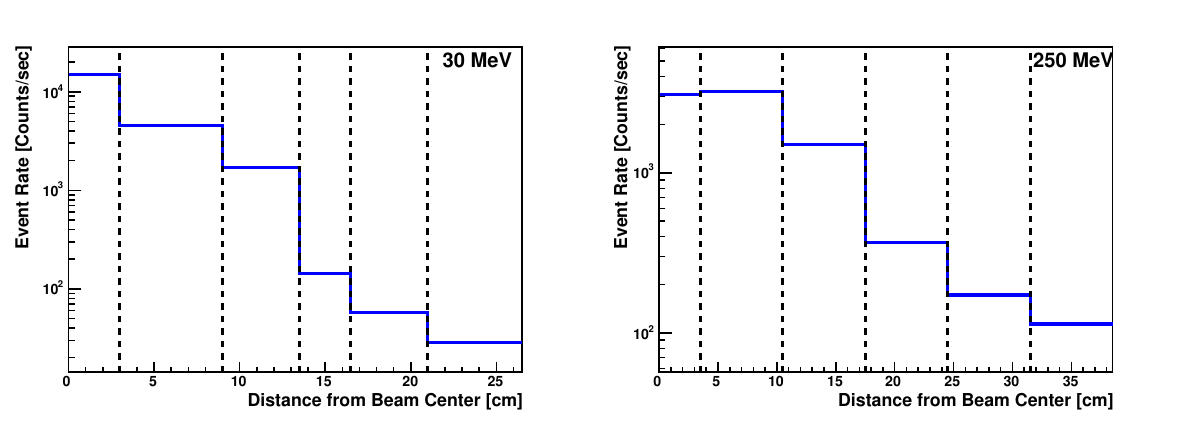}	
    \caption{Neutron beam profile at each distance from the beam center in the 30~MeV (left) and 250~MeV experiment (right). The event counting is performed for the peak region, 18--30 MeV for the 30 MeV experiment and 235--270 MeV for the 250~MeV experiment.}
    \label{fig:beam_profile}
\end{figure}

Figure~\ref{fig:beam_profile} shows the neutron-like event rates measured using BC-501A at some distances horizontally away from the beam center in the 30~MeV and 250~MeV experiments. 
The measurement method follows the one outlined in Section~\ref{sec:neutron_flux}.
From this measurement, we decided to place the HPGe detector at 24~cm away from the beam center, where the neutron event rate is reduced by an order of $\sim$3 and $\sim$1 compared to the center in the 30~MeV and 250~MeV experiment, respectively. 
A wider beam profile in the 250~MeV experiment is because our setup is made at more downstream as explained in Section~\ref{sec:experiment}. 
We also placed the CsI(Tl) scintillator at the same distance from the beam center as the HPGe detector. 

\subsection{The HPGe waveform analysis}

\begin{figure}[htbp]
\centering
    \includegraphics[width=0.49\textwidth]{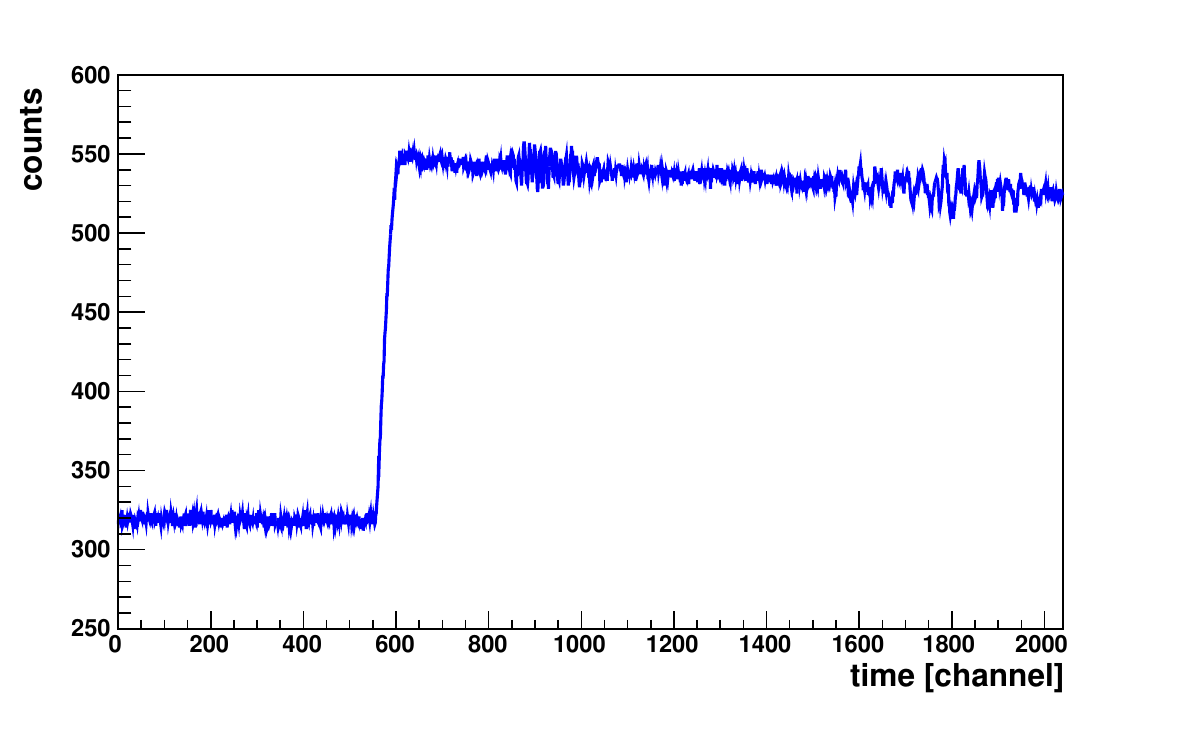}	
    \includegraphics[width=0.49\textwidth]{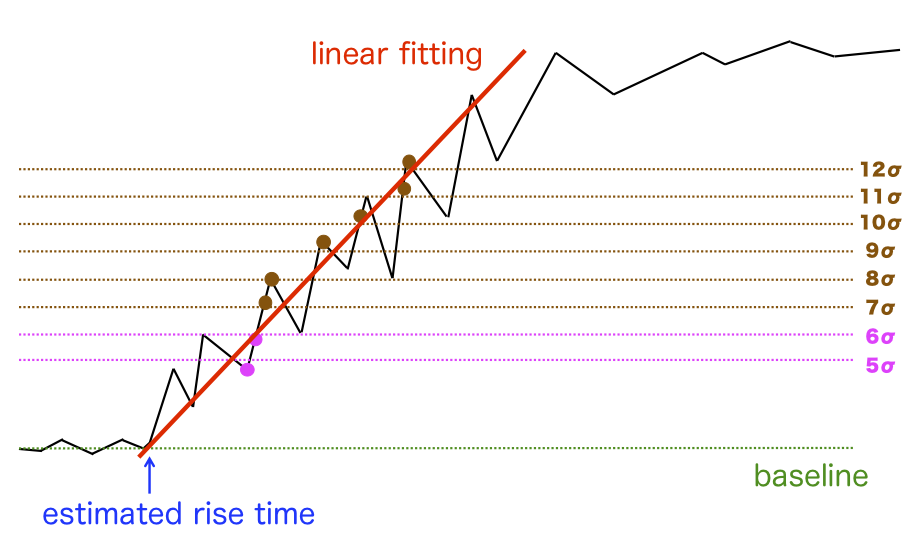}
    \caption{An example waveform from the HPGe detector (left) and a schematic drawing of the analysis scheme (right). In the left panel, 1 channel corresponds to 4~ns.}
    \label{fig:wf_ge}
\end{figure}

We designed an HPGe waveform analysis to reconstruct the time and energy for each detected event. 
Figure~\ref{fig:wf_ge} shows an example HPGe waveform taken by the digitizer. 
The baseline is calculated for each waveform as an average of the ADC (analog-to-digital converter) counts for the range of 0 to 300~channels, where 1 channel corresponds to 4~ns in time. 
Here, the standard deviation ($\sigma$) is computed together and used for determining the detected time. 
An event time is reconstructed as an intersection of the calculated baseline and the waveform rise edge which is obtained from a linear fitting to the points of 5$\sigma$ to 12$\sigma$ away from the baseline, as schematically shown in Figure~\ref{fig:wf_ge}. 
For the energy reconstruction, we use the maximum height from the baseline in each waveform, which is justified by the fact that the pre-amplifier output from HPGe is proportional to the waveform size. 

The defined physics variables were calibrated before each 30 and 250~MeV experiment on its energy and timing measurements. 
For the energy calibration, we made use of $\gamma$-rays from various radioactive isotopes, including $^{2}\rm{H}$, $^{60}\rm{Co}$, $^{40}\rm{K}$, $^{241}\rm{Am/Be}$, and $^{56}\rm{Fe}$, which enabled us to calibrate the detector up to $\sim$8~MeV with making sure its absolute scale within a 1\% level. 
The energy scale was monitored by performing a quick calibration with a few sources during the experiments and no significant shift was observed. 
We also calibrated the time response of the HPGe detector within a few ns uncertainty using two $\gamma$-rays emitted at the same time from a $^{60}\rm{Co}$ source with tagging one using a fast response detector, ${\rm LaBr_{3}(Ce)}$, and the other by HPGe

\subsection{Analysis time window}

\begin{figure}[h]
		\centering
		\includegraphics[width=1.0\textwidth]{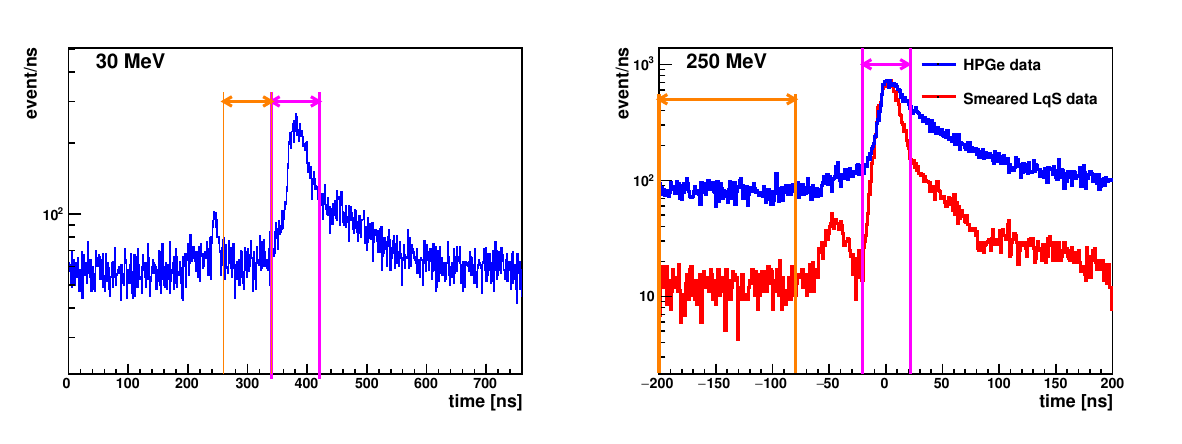}	
		\caption{ToF distributions of the HPGe detector (blue) in the 30~MeV (left) and 250~MeV experiment (right). In the right panel, the smeared ToF distribution of the BC-501A detector (red) is also shown. The on-timing and off-timing windows for each experiment are displayed as the arrows in magenta and orange, respectively.}
        \label{fig:HPGe_tof}
    \end{figure}

We performed the ToF analysis for the HPGe detector similarly to the BC-501A data as described in Section~\ref{sec:neutron_flux}. 
The resulting ToF distributions are shown in Figure~\ref{fig:HPGe_tof}. 
In the distribution from the 30~MeV experiment, a clear peak by the flash $\gamma$-ray is observed around 250~ns; however, it is not clearly observed in the 250~MeV experiment. 
This is perhaps because the background amount is expected to be more as is implied by the wider neutron beam profile in Figure~\ref{fig:beam_profile} and the higher energy beam. 
Therefore, we decide to utilize the BC-501A data to find the flash $\gamma$-ray point which is the baseline point for kinetic energy reconstruction. 
For this, we smeared the BC-501A's ToF distribution by resolution (standard deviation) difference in the time reconstruction between HPGe (6.8~ns) and BC-501A (1.9~ns). 
The smeared ToF distribution of the BC-501A detector is also shown in the right panel of Figure~\ref{fig:HPGe_tof}. 

We define two time windows, the on- and off-timing window, based on the reconstructed ToF distributions, which are used later for obtaining the signal and background energy spectra.
Each time window is indicated in Figure~\ref{fig:HPGe_tof}. 
The on-timing window corresponds to the range of 18--30~MeV and 235--270~MeV in neutron energies in the 30 and 250~MeV experiment, respectively. 
The off-timing window is defined using the ToF range where neutron rate is expected to be low. 
It is the range between the flash $\gamma$-ray peak and the bulk neutrons in the 30~MeV experiment, and defined as the range preceding the expected flash $\gamma$-ray position in the 250~MeV experiment, as indicated in Figure~\ref{fig:HPGe_tof}. 

\subsection{Observed $\gamma$-ray peaks}

\begin{figure}[htbp]
    \centering
    \includegraphics[width=0.9\textwidth]{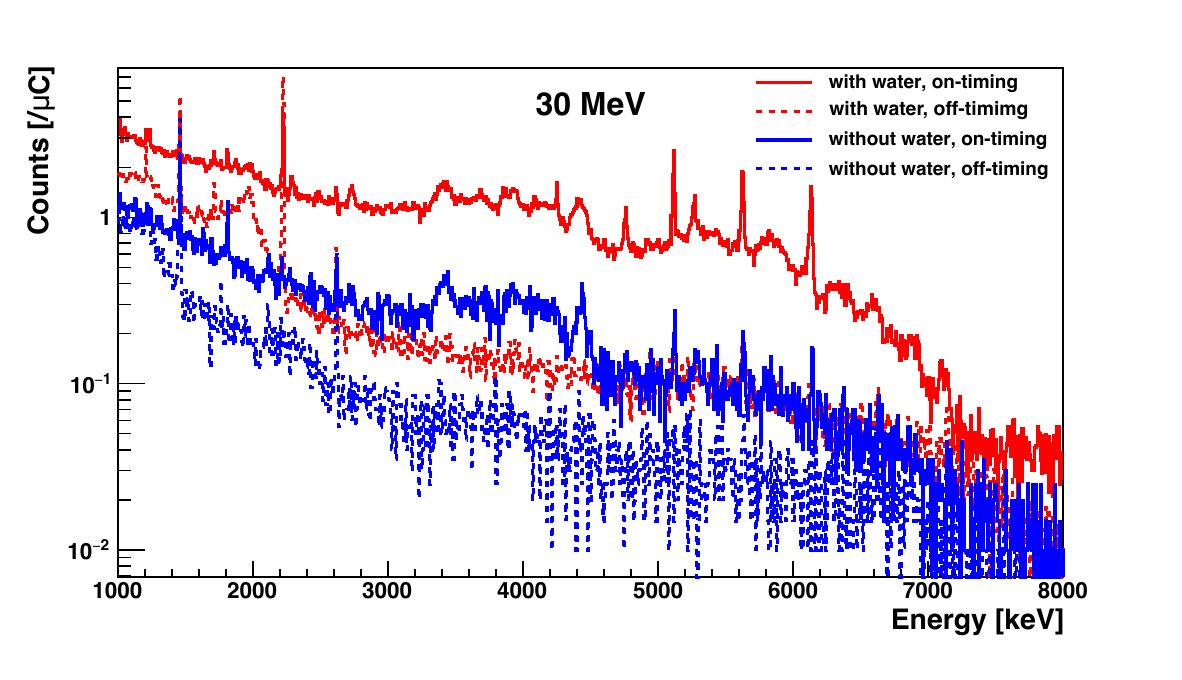}	
    \includegraphics[width=0.9\textwidth]{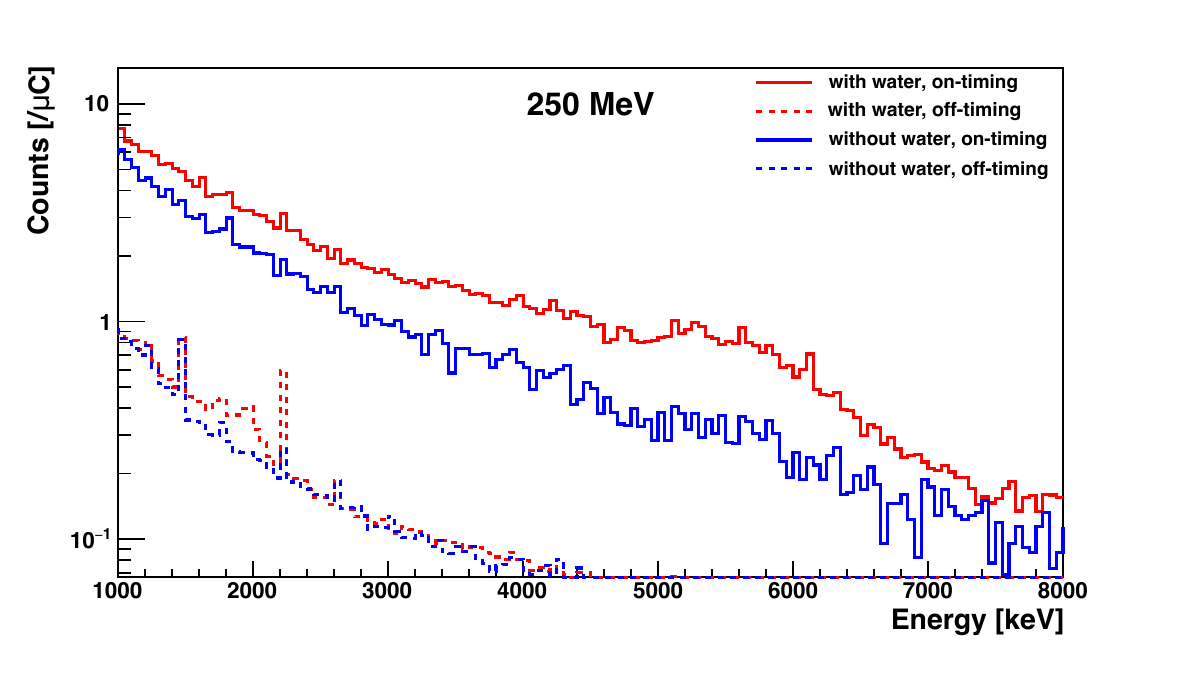}	
    \caption{Observed energy distributions in two configurations on the acrylic container(s) and for two timing windows from the 30~MeV (top) and 250~MeV experiment (bottom).}
    \label{fig:spectrum_w_wo_water_ToF}
    \end{figure}

Figure~\ref{fig:spectrum_w_wo_water_ToF} shows the measured energy spectra in two configurations and for the on- and off-timing windows from the 30~MeV and 250~MeV experiments. 
Note that the spectra for the off-timing window is normalized to the size of the on-timing window. 
Multiple $\gamma$-ray peaks are observed in the water-filled configuration. 
The observed peaks, their parent nuclei and likely physics processes are summarized in Table~\ref{tab:observed_gamma}. 
Many of them were observed also in the previous experiment (E487) at $E_{n} = 80$~MeV~\cite{2024PhRvC.109a4620A}.
The $\gamma$-rays of 7.12, 6.92, 6.13, and 2.74~MeV are expected from different excited states of ${\rm ^{16}O}$ and clearly observed in both experiments. 
The parent process for these $\gamma$-rays, the $(n,n^{\prime})$ scattering, does not cause an additional nucleon in the final state. 
The 6.32, 5.27, and 2.30~MeV $\gamma$-rays are emitted from excited states of ${\rm ^{15}N}$, where a proton is released from the $(n,np)$ scattering. 
Here, the final state neutron and proton may form a bound state of a deuteron, but our measurement is not able to tell this. 
It should also be note that the proton release possibly occurs after the $(n,n^{\prime})$ scattering via ${\rm ^{16}O^{*}} \to {\rm ^{15}N^{*}} + p$ and again the current experiment is not sensitive enough to distinguish this. 
Similarly, the 6.18 and 5.18~MeV $\gamma$-rays are produced from excited states of ${\rm ^{15}O}$ with an additional neutron emission. 
The 5.10, 4.91, and 2.31~MeV $\gamma$-rays are expected from excited states of ${\rm ^{14}N}$. 
The 5.10 and 4.91~MeV peaks are not observed very clearly, but they are considered in the later spectral fitting. 
The 2.31~MeV peak cannot be distinguished from the 2.30~MeV $\gamma$-ray peak by the HPGe detector, hence we measure the inclusive cross section for these two $\gamma$-rays. 
Intense spectral peaks are formed by the 3.68~MeV $\gamma$-ray from ${\rm ^{13}C}(\frac{3}{2}^{-})$ and the 4.44~MeV $\gamma$-ray from ${\rm ^{12}C}(2^{+})$. 
Many peaks originating from thermal neutron capture process are also observed: 2.22~MeV from ${\rm ^{2}H}$, 3.84~MeV from ${\rm ^{17}O}$, and 7.63~MeV from ${\rm ^{57}Fe}$. 
In addition, the 1.46~MeV $\gamma$-ray from ${\rm ^{40}K}$ and the 2.61~MeV $\gamma$-ray from ${\rm ^{208}Tl}$ are produced in the highly-radiated environment. 
These $\gamma$-rays are not the measurement targets of the present work. 

    \begin{table}[htbp]
    \centering
    \caption{Summary of the neutron-oxygen reaction-induced $\gamma$-rays identified by the HPGe detector with their parent excited nuclei and physics processes.}
    \begin{tabular}{c c l}
    \hline \hline
        Energy [MeV] & Parent nucleus ($J^\pi$) & Physics process \\ \hline \hline
        2.30 & ${\rm ^{15}N}(\frac{7}{2}^{+})$ & ${\rm ^{16}O}(n,np){\rm ^{15}N^{*}}$ \\
        2.31 & ${\rm ^{14}N}(0^{+})$ & ${\rm ^{16}O}(n,2np){\rm ^{14}N^{*}}$ \\ 
        2.74 & ${\rm ^{16}O}(2^{-})$ & ${\rm ^{16}O}(n,n^{\prime}){\rm ^{16}O^{*}}$ \\ 
        3.68 & ${\rm ^{13}C}(\frac{3}{2}^{-})$ & ${\rm ^{16}O}(n,\alpha){\rm ^{13}C^{*}}$ \\
        4.44 & ${\rm ^{12}C}(2^{+})$ & ${\rm ^{16}O}(n,n\alpha){\rm ^{12}C^{*}}$ \\ 
        4.91 & ${\rm ^{14}N}(0^{-})$ & ${\rm ^{16}O}(n,2np){\rm ^{14}N^{*}}$ \\ 
        5.10 & ${\rm ^{14}N}(2^{-})$ & ${\rm ^{16}O}(n,2np){\rm ^{14}N^{*}}$ \\ 
        5.18 & ${\rm ^{15}O}(\frac{1}{2}^{+})$ & ${\rm ^{16}O}(n,2n){\rm ^{15}O^{*}}$ \\ 
        5.27 & ${\rm ^{15}N}(\frac{5}{2}^{+})$ & ${\rm ^{16}O}(n,np){\rm ^{15}N^{*}}$ \\ 
        6.13 & ${\rm ^{16}O}(3^{-})$ & ${\rm ^{16}O}(n,n^{\prime}){\rm ^{16}O^{*}}$ \\ 
        6.18 & ${\rm ^{15}O}(\frac{3}{2}^{-})$ & ${\rm ^{16}O}(n,2n){\rm ^{15}O^{*}}$ \\ 
        6.32 & ${\rm ^{15}N}(\frac{3}{2}^{-})$ & ${\rm ^{16}O}(n,np){\rm ^{15}N^{*}}$ \\
        6.92 & ${\rm ^{16}O}(2^{+})$ & ${\rm ^{16}O}(n,n^{\prime}){\rm ^{16}O^{*}}$ \\
        7.12 & ${\rm ^{16}O}(1^{-})$ & ${\rm ^{16}O}(n,n^{\prime}){\rm ^{16}O^{*}}$ \\ \hline \hline
    \end{tabular}
    \label{tab:observed_gamma}
    \end{table}

\subsection{Background estimation} \label{sec:bkgestimate}

For the $\gamma$-ray production cross section measurement to be performed later, we estimate background events to extract the intensity of $\gamma$ signals. 
In this study, four types of background events are considered as following the previous experiment~\cite{2024PhRvC.109a4620A}: fast neutron hits, non-water background, time-uniformly distributed $\gamma$ and $\beta$ background, and scattered neutron-induced events. 
Each of them is described in the following part. 

\subsubsection*{Fast neutrons directly entering the detector} \ \   
Neutrons which are scattered by the acrylic container(s) or incident off beam axis and enter the HPGe directly can be background events.
For the measurement of this background, a CsI(Tl) scintillator was placed at upstream of the acrylic container in a symmetric position with respect to HPGe across the beam axis. 
Based on the PSD analysis on CsI(Tl) data, which has been established in the previous tests~\cite{2018PTEP.2018d3H01A,2024PhRvC.109a4620A}, such background is negligible up to $\sim$7~MeV and accounts for 2\% of the observed $\gamma$-ray intensity in the 7--8~MeV region. 

\subsubsection*{Non-water background} \ \  
$\gamma$-rays produced from sources other than water, such as the acrylic container(s) and experimental stage, constitute background.
This can be subtracted by the measurement with an empty configuration, as shown as the blue histogram in Figure~\ref{fig:spectrum_w_wo_water_ToF}. 

\subsubsection*{$\gamma$-rays from thermal neutron capture and $\beta$'s from beta decay} \ \
$\gamma$-rays originating from thermal neutron capture and $\beta$'s from beta decay are expected to distribute uniformly in time as their emission timescale is much longer than the beam repetition cycle. 
Therefore, we estimate these background events by the off-timing spectrum created with a TOF cut (see Figure~\ref{fig:spectrum_w_wo_water_ToF}). 

\subsubsection*{Scattered fast neutron-induced $\gamma$-rays} \ \ 
There still remains continuous background in the observed HPGe spectrum, which are expected to arise from reactions between neutrons scattered off water and the surrounding materials. 
Such background was confirmed in E487~\cite{2024PhRvC.109a4620A} and its estimation method was established. 
Therefore, we take the same estimation method in the current study, which will be described as a continuous background template in Section~\ref{sec:cross_section}.

\section{Cross Section Analysis} \label{sec:cross_section}

In this section, we describe the ${}^{16}\mathrm{O}(n, \mathrm{X}\gamma)$ cross section measurement for the peaks shown in Table~\ref{tab:observed_gamma}. 
We take a binned fitting method on the energy spectrum obtained after some background subtraction. 
First, we subtract the non-water background and the uniform-in-time background (see Section~\ref{sec:bkgestimate}) as,  
    \begin{equation}
        N^\mathrm{Data}_{\mathrm{bin}} = \left( N^\mathrm{Water, ON}_{\mathrm{bin}} - N^\mathrm{Water, OFF}_{\mathrm{bin}} \right) - \left( N^\mathrm{Empty, ON}_{\mathrm{bin}} - N^\mathrm{Empty, OFF}_{\mathrm{bin}} \right),
    \end{equation}
where $N^\mathrm{Water, ON}_{\mathrm{bin}}$ is the number of events in the bin-th energy bin of the observed spectrum from the water-filled container data in the on-timing window. 
Accordingly, $N^\mathrm{Water, OFF}_{\mathrm{bin}}$ corresponds to that from the water-filled container data in the off-timing window.
$N^\mathrm{Empty, ON}_{\mathrm{bin}}$ and $N^\mathrm{Empty, OFF}_{\mathrm{bin}}$ correspond to the empty-container data for the on- and off-timing window, respectively. 
We then perform the fitting to this spectrum after these subtraction with including the remaining continuous background as a part.

\subsection{Signal and background templates}

Spectrum templates of $\gamma$-rays from discrete levels of excited nuclei are produced by a Monte Carlo simulation based on the GEANT4 package~\cite{bib:geant4} in order to include a geometrical effect of the setup on the spectra observed in the HPGe detector.
The simulated $\gamma$-rays are listed in Table~\ref{tab:observed_gamma}.
The $\gamma$-rays with isotropic momentum direction are generated at a point following an emission point probability density function (PDF) in the water sample.
The emission point PDF was generated based on the position where neutron inelastic scattering occurred in the MC simulation shooting neutron with the measured beam profile (Fig.~\ref{fig:beam_profile}) onto the water target.
There is almost no change in the spectrum template if the PDF is replaced with a uniform distribution. Thus, a systematic uncertainty due to the emission point is negligibly small.

The Doppler broadening effect is taken into consideration for the states with a shorter lifetime than their interaction time scale. 
Thus, we apply a 1\% smearing with a Gaussian distribution for the 2.30, 2.74, 3.68, 4.91, 5.18, 6.18, 6.32, 6.92 and 7.12~MeV $\gamma$-rays, and gives a 2\% smearing for the 4.44~MeV $\gamma$-ray, respectively.
Because of the Doppler broadening and the detector resolution, it is difficult to separate contributions from the 2.30~MeV and 2.31~MeV $\gamma$-rays. 
Thus, the 2.30~MeV $\gamma$-ray template is included in the fit for the inclusive estimate of these two peaks. 

The simulated deposited energy in the HPGe detector is smeared with an energy resolution function. 
It is determined by fitting energy resolution at calibration points with a function $\sigma/E = p_0 + p_1/E$ ($\sigma/E = p_0 + p_1/E^2$) for the 30 (250)~MeV experiment, where $\sigma$ is the standard deviation determined by a Gaussian fit and $E$ is the peak energy. As a result of the curve fitting, the parameters, ($p_0, p_1$), are determined as ($(3.51 \pm 0.03)\times 10^{-1}, (201.5 \pm 2.8)\times 10^{3}$) for the 30 MeV experiment and ($(2.32 \pm 0.10)\times 10^{-1}, (5.38 \pm 0.20)\times 10^{2}$) for the 250~MeV experiment, respectively.
An effect of the fitting errors on the cross section estimation is negligible.

An uncertainty about the $\gamma$-ray detection efficiency at the HPGe detector is estimated based on the calibration data in which a ${}^{60}$Co source was placed at 6, 9 and 12~cm away from the HPGe detector. Simulated spectra in the identical setup with the calibration were compared with the calibration data. We found that the simulated spectra reproduced the observed data within a 10\% difference at each source position. Thus, this 10\% is taken into consideration as a systematic uncertainty in the spectrum fitting.

As discussed above, the continuous background spectrum formed by $\gamma$-rays from the scattered neutron reactions on surrounding materials is considered in the fit. 
Its spectrum shape is estimated based on the on-timing data with the empty acrylic container(s). 
The template as a smooth function is obtained by fitting the empty sample data with an exponential function $f(E) = \exp(p_0 + p_1 E)$, where $E$ is the observed energy at the HPGe detector in the unit of keV. 
The determined parameters are $p_0 = 8.67 \pm 0.06$ and $p_1 = (-5.52 \pm 0.15) \times 10^{-4}$ for the 30~MeV experiment, and $p_0 = 9.65 \pm 0.05$ and $p_1 = (-5.27 \pm 0.12) \times 10^{-4}$ for the 250~MeV experiment. 
The obtained spectra from two experiments are consistent to each other within 10\%.

\subsection{Least chi squared fit}

\begin{figure}[h]
    \centering
    \includegraphics[width=1.0\textwidth]{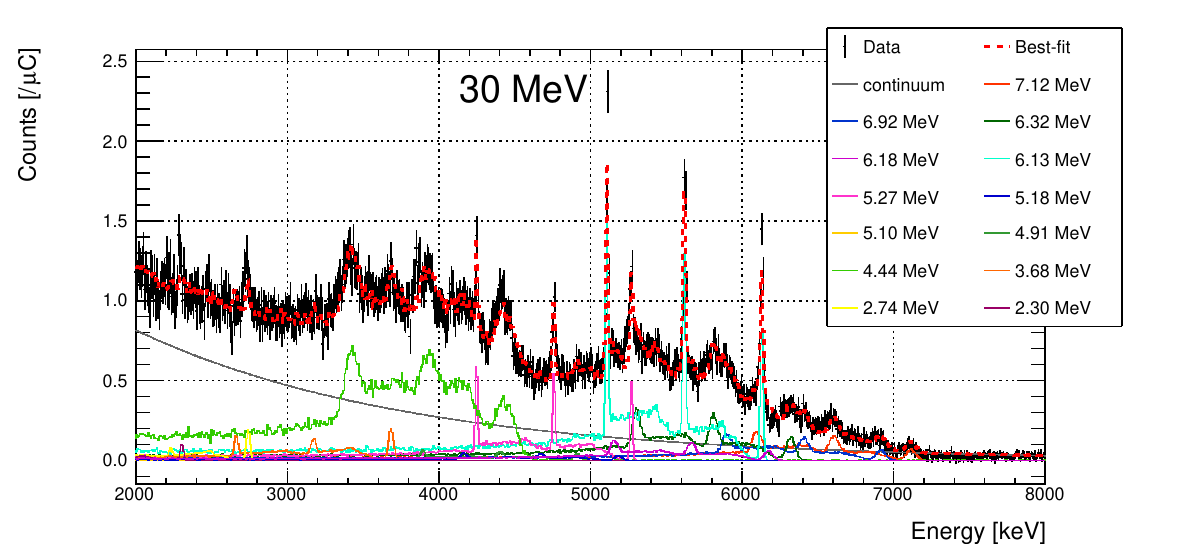}
    \includegraphics[width=1.0\textwidth]{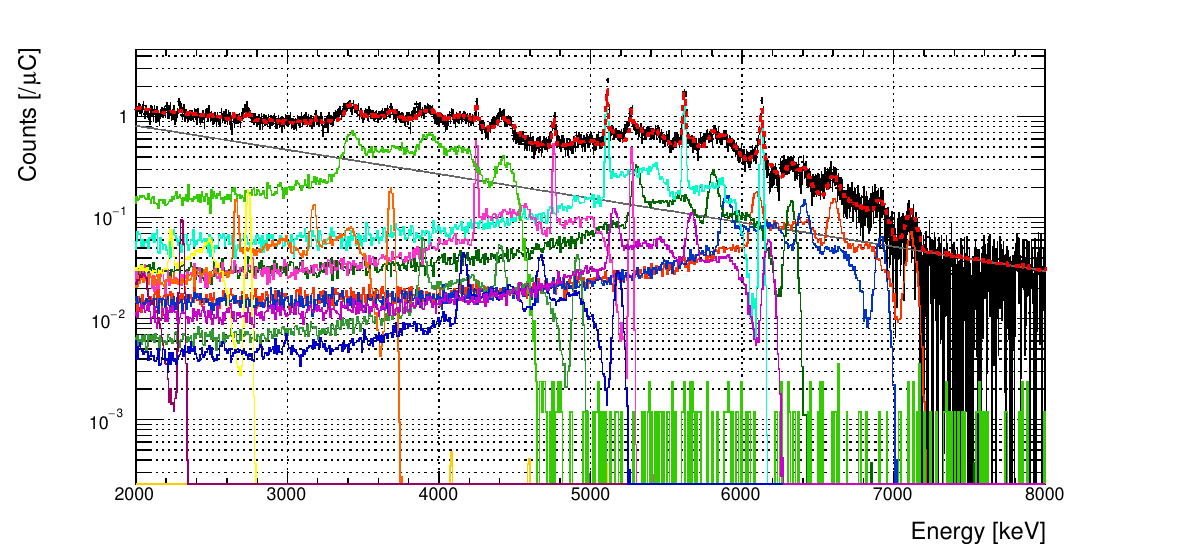}
    \caption{Energy spectra of the signal $\gamma$-ray in the 30 MeV experiment (black points) overlaid with the best-fit of the spectrum fitting (dashed red line) in a linear scale (top) and a logarithmic scale (bottom). The signal and background templates are overlaid to show the result of fit, which are colored in as follows: orange (7.12~MeV), azure (6.92~MeV), deep green (6.32~MeV), magenta (6.18~MeV), light blue (6.13~MeV), pink (5.27~MeV), blue (5.18~MeV), dark yellow (5.10~MeV), green (4.91~MeV), light green (4.44~MeV), light orange (3.68~MeV), yellow (2.74~MeV), purple (2.30/2.31~MeV) and gray (background continuum).}
    \label{fig:best_fit_30MeV}
\end{figure}

\begin{figure}[h]
    \centering
    \includegraphics[width=1.0\textwidth]{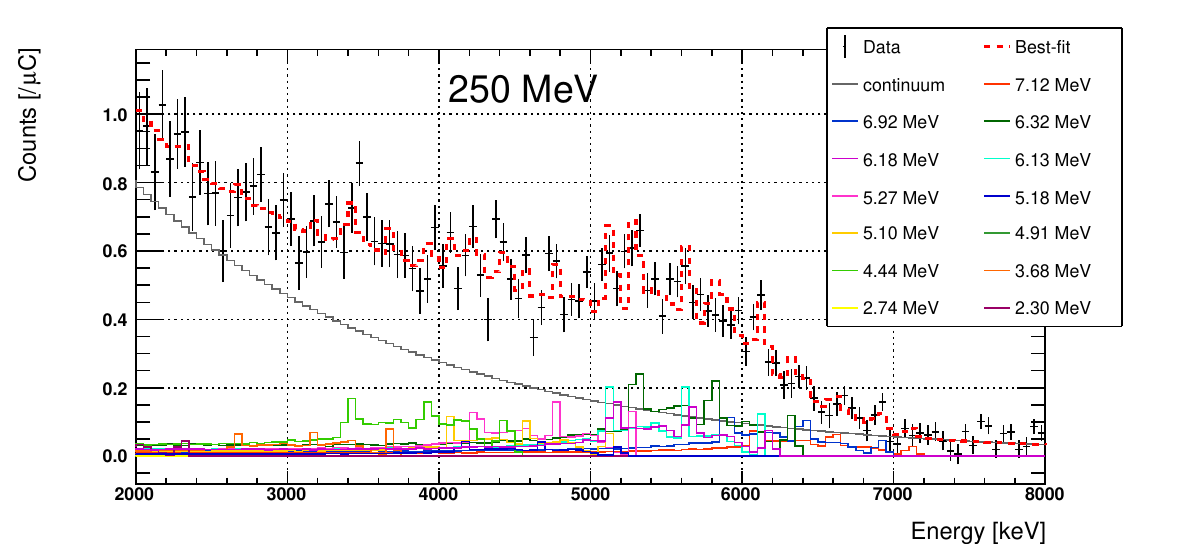}
    \includegraphics[width=1.0\textwidth]{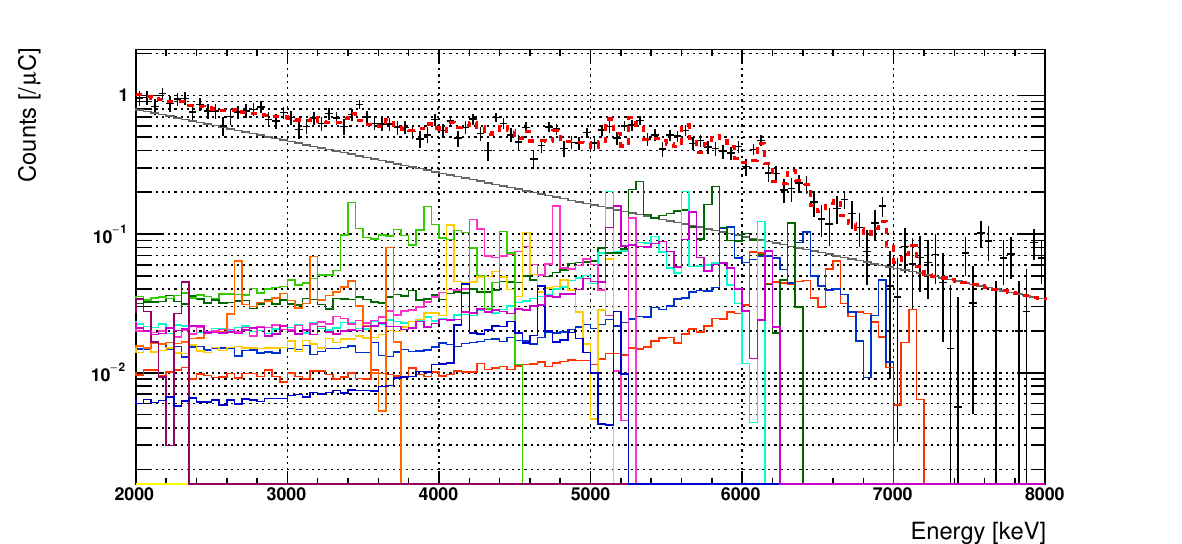}
    \caption{Energy spectra of the signal $\gamma$-ray in the 250~MeV experiment (black points) overlaid with the best-fit of the spectrum fitting (dashed red line) in a linear scale (top) and a logarithmic scale (bottom). The color notation is the same as Figure~\ref{fig:best_fit_30MeV}.}
    \label{fig:best_fit_250MeV}
\end{figure}

The data spectrum is fit with the simulated templates based on a chi-squared taking the systematic uncertainty into consideration. The definition of the chi-squared is denoted as, 
\begin{equation}
\begin{split}
     \chi^{2} &= \sum_{\mathrm{bin}} \left( \frac{N^\mathrm{Data}_{\mathrm{bin}} - \eta \times N^\mathrm{MC}_{\mathrm{bin}}}{\sigma^{\mathrm{stat}}_{\mathrm{bin}}} \right)^2 + \left( \frac{1 - \eta}{\sigma^{\mathrm{syst}}} \right)^2, \\
     & N^\mathrm{MC}_{\mathrm{bin}} = \sum_j f_j \times n_{\mathrm{bin}}^{j},
\end{split}
\end{equation}
where $\eta$ is a nuisance parameter characterizing the systematic uncertainty from the $\gamma$-ray detection efficiency at the HPGe detector. As mentioned above, we assigned a 10\% uncertainty on the detection efficiency.
$N^\mathrm{Data}_{\mathrm{bin}}$ and $N^\mathrm{MC}_{\mathrm{bin}}$ represent the number of the observed events at each energy bin for the observed data and the Monte Carlo simulation, respectively.
The Monte Carlo prediction, $N^\mathrm{MC}_{\mathrm{bin}}$, consists of the summation of the spectrum templates ($n^{j}_{\mathrm{bin}}$) with scale factors ($f_{j}$) where $j$ corresponds to the template number, 0: continuous background, 1: 2.30 \& 2.31~MeV, 2: 2.74~MeV, 3: 3.68~MeV, 4: 4.44~MeV, 5: 4.91~MeV, 6: 5.10~MeV, 7: 5.18~MeV, 8: 5.27~MeV, 9: 6.13~MeV, 10: 6.18~MeV, 11: 6.32~MeV, 12: 6.92~MeV, and 13: 7.12~MeV. 
The statistical errors of the observed data and simulation are considered in $\sigma^{\mathrm{stat}}_{\mathrm{bin}}$ at each energy bin.
Minimization of the chi-squared and error estimation were performed using a dedicated tool, MINUIT~\cite{James:1975dr}, with the scale factors and the nuisance parameter as fitting parameters.
The best-fit spectra are shown in Figures~\ref{fig:best_fit_30MeV} and \ref{fig:best_fit_250MeV}, respectively.
We show the best-fit results with $\pm 1\sigma$, but only give the 90\% upper limits for some components whose scale factor is zero-consistent as a result of the fit.
The minimum chi-squared value is 564.2 (135.7) in the degree of freedom 635 (89) for the 30 (250)~MeV case.
Therefore, the assumption that the observed spectra consist of $\gamma$-rays from discrete levels of nuclei and the continuum component works well in terms of the goodness of fit.

\subsection{$\gamma$-ray production cross section}
The $\gamma$-ray production cross section at each discrete level can be computed as, 
\begin{equation}
     \sigma_{j}^{\gamma} = f_{j} \times \frac{N^{\mathrm{MC}}_{\mathrm{gen}}}{T_{\mathrm{Oxygen}} \times \phi_{n}},
\end{equation}
where $N^{\mathrm{MC}}_{\mathrm{gen}}$ is the number of generated $\gamma$-rays in producing the spectrum template, $T_{\mathrm{Oxygen}}$ is the number of ${}^{16}$O existing in the water sample, and $\phi_{n}$ is the neutron flux. 
The numbers used here are as follows: $N^{\mathrm{MC}}_{\mathrm{gen}} = 1.0 \times 10^{8}$ ($1.0 \times 10^{8}$), $T_{\mathrm{Oxygen}} = 2.3 \times 10^{26}$ ($4.6 \times 10^{26}$), $\phi_{n} = 1.4 \times 10^{4}$ ($1.2 \times 10^{3}$) neutrons/cm${}^{2}$/$\mu$C for the 30 (250)~MeV experiment.
Table~\ref{table:xs_result} shows the summary of the measured $\gamma$-ray production cross sections in the E525 experiment.
In addition to the error from the fit, the neutron flux uncertainty is propagated for the total uncertainty on the cross section result (see Table~\ref{table:flux_errors}).
As mentioned above, we give the 90\% upper limit for the peaks whose statistics was not significant to be measured in the fit.
The total $\gamma$-ray production cross sections are measured as $\sigma_{\mathrm{total}} = 112.5 \pm 7.6$ at 30~MeV and $\sigma_{\mathrm{total}} = 74.3 \pm 11.1$ at 250~MeV by integrating over the 2 to 8~MeV $\gamma$-rays and subtracting the continuum component.
Discussion about the results is given in the next section.

\begin{table}[t]
    \centering
    \caption{The $\gamma$-ray production cross section results from the 30 MeV and 250 MeV experiments. For the 2.30 and 2.31~MeV $\gamma$-rays, the inclusive measurement results are shown.}
    \begin{tabular}{cccc}
    \hline \hline
     No.  & Peak Energy [MeV] & Cross Section (30~MeV)~[mb] & Cross Section (250~MeV)~[mb] \\ \hline \hline
        1 & 2.30/2.31         &  $0.8 \pm 0.7$              &  $<6.9$ (90\% C.L.)          \\
        2 & 2.74              &  $2.2 \pm 0.9$              &  $<2.8$ (90\% C.L.)          \\
        3 & 3.68              &  $4.1 \pm 1.2$              &  $3.7 \pm 2.6$               \\
        4 & 4.44              & $37.5 \pm 6.8$              & $10.8 \pm 3.3$               \\
        5 & 4.91              &  $1.9 \pm 1.2$              &  $<3.7$ (90\% C.L.)          \\
        6 & 5.10              &  $<0.1$ (90\% C.L.)         &  $5.4 \pm 2.6$               \\
        7 & 5.18              &  $1.3 \pm 0.9$              &  $<6.2$ (90\% C.L.)          \\
        8 & 5.27              &  $9.9 \pm 1.9$              &  $8.0 \pm 2.7$               \\
        9 & 6.13              & $22.3 \pm 4.1$              &  $9.4 \pm 3.1$               \\
       10 & 6.18              &  $4.6 \pm 1.3$              &  $9.2 \pm 3.4$               \\
       11 & 6.32              & $12.9 \pm 2.5$              & $15.3 \pm 4.2$               \\
       12 & 6.92              &  $6.0 \pm 1.5$              &  $7.2 \pm 2.6$               \\
       13 & 7.12              &  $9.0 \pm 1.9$              &  $5.3 \pm 2.3$               \\ \hline \hline
    \end{tabular}
    \label{table:xs_result}
\end{table}

\section{Discussion} \label{sec:discussion}

Ref.~\cite{bib:nelson} reported the measurement results of the ${}^{16}\mathrm{O}(n, \mathrm{X}\gamma)$ cross sections using a white neutron beam which has a broad kinetic energy range up to 200~MeV. 
This work gives the unique results using nearly mono energy neutrons at 30 and 250 MeV, and the cross sections at 250~MeV are the first measurement results above 200~MeV. 
In addition, more types of $\gamma$-rays are identified in this work than the work done in Ref.~\cite{2024PhRvC.109a4620A} because of the improvement on the energy resolution thanks to the HPGe detector. For example, the $\gamma$-rays of 2.74~MeV from $\rm {}^{16}O(2^-)$, 4.91~MeV from $\rm {}^{14}N(0^-)$, 5.10~MeV from $\rm {}^{14}N(2^-)$, and 6.18~MeV from $\rm {}^{15}N(\frac{3}{2}^{-})$ are newly measured.
It is difficult to identify the 5.10~MeV $\gamma$-ray based on its photopeak in Ref.~\cite{bib:nelson} suffering from the double escape peak of the intense 6.13~MeV $\gamma$-ray.
However, our spectrum fit method improves sensitivity to such peaks due to information of spectrum shape, e.g., the single and double escape peaks and Compton edge.

In the present work, the 6.18~MeV and 6.32~MeV $\gamma$-rays are observed and their cross sections are measured. 
They are due to deexcitation from a $\frac{3}{2}$-hole state of $\rm {}^{15}O$ and $\rm {}^{15}N$, respectively, which are expected to be generated as a result of the NCQE reaction by neutrinos as well~\cite{2012PhRvL.108e2505A,2013PhRvD..88i3004A}.
The measured cross section of the 6.32~MeV $\gamma$-ray is larger than that of the 6.18~MeV $\gamma$-ray both at 30 and 250~MeV. 
This is considered because a neutron is more likely to collide with a proton than a neutron due to the nucleon isospin. 

\begin{figure}[h]
    \centering
    \includegraphics[width=1.0\textwidth]{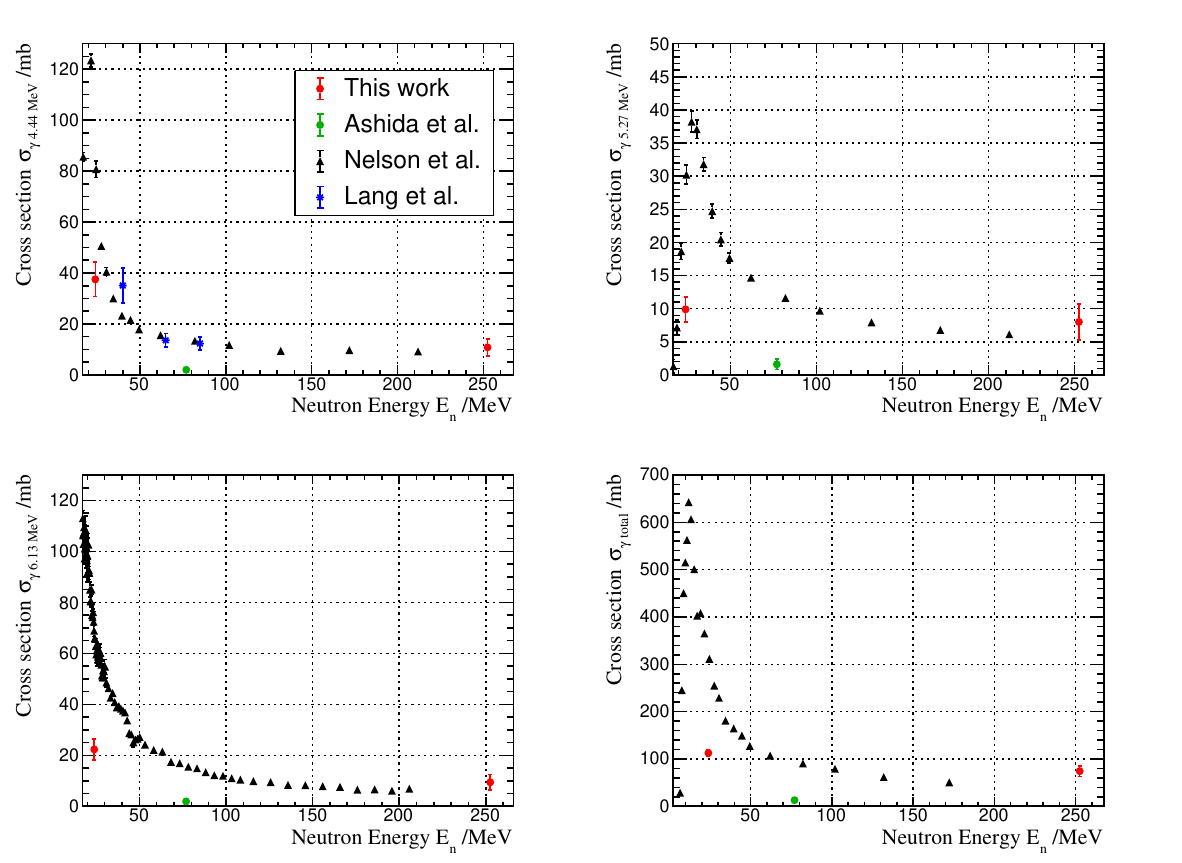}	
    \caption{Measured cross sections of the 4.44~MeV (top left), 5.27~MeV (top right), 6.13~MeV (bottom left) and total (bottom right) $\gamma$-ray production as a function of neutron kinetic energy. The result of this work (red points) and E487 (green point) are compared with Ref.~\cite{bib:nelson} (black) in each plot. The 6.13~MeV $\gamma$-ray production by proton-oxygen interaction~\cite{PhysRevC.35.1214} (blue points) is displayed as well.}
    \label{fig:comp_nelson}
\end{figure}

Figure~\ref{fig:comp_nelson} shows comparison between measurements of the ${}^{16}\mathrm{O}(n, \mathrm{X}\gamma)$ cross sections of the 4.44~MeV, 5.27~MeV, 6.13~MeV peaks and the total $\gamma$-ray production from the present work and Refs.~\cite{bib:nelson,2024PhRvC.109a4620A}.
The measured cross section at a 250~MeV neutron energy shows consistency with the adjacent data points from Ref.~\cite{bib:nelson} within the estimated uncertainty.
In contrast, our measurement results at a 30~MeV neutron energy show smaller cross sections by a factor of 2 to 3 than Ref.~\cite{bib:nelson}. 
This is under investigation in more detail using a Monte Carlo simulation, but no longer discussed here as it is beyond the current scope. 

In this work, as shown in Tables~\ref{tab:observed_gamma} and \ref{table:xs_result}, the measured cross sections for the decay mode with neutron emission ($\gamma$-rays of 2.31, 4.91, 5.10, 5.18, and 6.18~MeV) are smaller compared to others. 
At water Cherenkov detectors, because protons and ions with energies of current interest are not observable, nuclear reactions with an additional neutron emission are more crucial. 
Our results may indicate that the secondary $\gamma$-ray emission is likely to be caused by the initial neutron from the primary neutrino interaction in many cases. 
In the NCQE measurement at T2K~\cite{2019PhRvD.100k2009A}, as mentioned in Section~\ref{sec:intro}, the observed number of events at large Cherenkov angles, where events with more $\gamma$-rays are reconstructed, was smaller than the expectation by Monte Carlo simulation in the neutrino-dominant sample. 
In contrast, this mismatch was found to be mitigated for the antineutrino-dominant sample. 
The kinetic energy of knocked-out nucleons by neutrino interactions is larger than that by antineutrinos and hence the observation in T2K implies that the current model predicts more secondary $\gamma$ emission for higher energy neutrons. 
This could be improved by the measured fact in this work that secondary $\gamma$-rays are dominantly produced by the initial neutron. 
Another useful reference is measurement of the neutron multiplicity for neutrino interactions performed by T2K using its charged-current (CC)-dominant sample~\cite{2019TAUP.Akutsu.T2K.Neutron,bib:rakutsu}. 
In this study, the observed number of neutrons after the CC interaction was significantly smaller than the predictions by three different neutrino interaction models. 
While the results are from the CC-dominant sample, the same secondary interaction model is used as NCQE in T2K. 
This study accesses the final state neutron directly than Ref.~\cite{2019PhRvD.100k2009A}, as it was done via $\gamma$-rays, and strongly supports the indication that the current model contributes to more secondary neutron interactions. 
In addition, in Ref.~\cite{2024PhRvD.109a1101S} using the atmospheric neutrino NCQE sample, the Cherenkov angle distributions from observed data and different secondary interaction models are compared. 
This disfavors the secondary interaction model based on intranuclear cascade model by Bertini~\cite{G4BERT_WRIGHT2015175} used in the two T2K studies above. 
Instead, the INCL++ model~\cite{INCL_PhysRevC.87.014606} is favored in this sample, and therefore comparing the prediction from this model and the current beam test results would be interesting.


\section{Conclusion} \label{sec:conclusion}
%
The E525 experiment to measure $\gamma$-rays from the neutron-oxygen interaction was carried out at RCNP using a quasi-mono energetic neutron beam at 30 and 250~MeV. 
A high energy resolution is achieved using the HPGe detector and helps to identify multiple $\gamma$-ray peaks.
Neutron flux measurement using the BC-501A scintillator confirms the absolute flux of quasi-mono energitic neutrons by lithium-proton interaction and its corresponding timing window for the signal $\gamma$-rays in the HPGe detector.
We gave a special care about the choice of proton beam current in order to avoid a potential human error on the current measurement suspected in the former E487 experiment.
The background estimation and subtraction were done based on the off-timing and the empty container data.
A spectrum fitting method is employed based on a chi-squared to measure the cross section for each peak and the remaining background events simultaneously.
The production cross section of thirteen kinds of $\gamma$-rays and the total of them were measured, and this work gives the first measurement results at a 250~MeV neutron energy.
The obtained data are useful for validation of neutron inelastic interaction models used in the detector simulation for neutrino event prediction. 
Comparing the simulation result with the present experimental data will help improving the systematic uncertainty caused by the neutron-oxygen interaction subsequently occurring after the neutrino NCQE interaction in water Cherenkov detectors.
	

    \section*{Acknowledgments}
    The authors are thankful to the RCNP staff for giving beam time to the experiment and the accelerator group for supplying stable beams. This work was supported by Japan MEXT KAKENHI Grant Number 17J06141, 18H05537, 26400292, and 25105002.


\bibliographystyle{unsrt}
\bibliography{reflist}

\end{document}